\documentstyle[aps,epsf,prl,multicol]{revtex}

\newenvironment{mytitle}{\begin{center} \large \bf }{\\ [.1in]\end{center}}
\newenvironment{myauthor}{\begin{center} }{\\ [.1in]\end{center}} 
\newenvironment{myinstit}{\begin{center} \it}{\end{center}}
\begin{document}
\thispagestyle{empty}

\begin{mytitle}
Periodic Chaotic Billiards: Quantum-Classical Correspondence in Energy Space
\end{mytitle}

\begin{myauthor}
G. A. Luna-Acosta,  J. A. M\'endez-Berm\'udez, and F. M. Izrailev
\end{myauthor}

\begin{myinstit}
Instituto de F\'{\i}sica, Universidad Aut\'onoma de Puebla,
Apdo. Postal J - 48, Puebla 72570,  Mexico.
\end{myinstit}

\begin{abstract}
We investigate the properties of eigenstates and local
density of states (LDOS) for a periodic 2D 
rippled billiard, focusing on their quantum-classical correspondence in
energy representation. To construct the classical counterparts of 
LDOS and the structure of eigenstates (SES), the effects of the boundary are 
first incorporated 
(via a canonical transformation) into an effective potential, 
rendering the one-particle motion in the 2D rippled billiard equivalent to 
that of two-interacting particles in 1D geometry. We show that classical
counterparts of SES and LDOS in the case of strong
chaotic motion reveal quite a good correspondence with the quantum quantities. 
We also show that the main features of the SES and LDOS can be 
explained in terms of the underlying classical dynamics, in particular 
of certain periodic orbits. 
On the other hand, statistical
properties of eigenstates and LDOS turn out to be
different from those prescribed by random matrix theory. 
We discuss the quantum effects responsible for the non-ergodic character
of the eigenstates and individual LDOS that seem to be generic for this
type of billiards with a large number of transverse channels.
\end{abstract}

\newpage

\section{Introduction}

The goal of this study is to deepen our understanding of the
quantum-classical correspondence for chaotic systems by analyzing
the properties of eigenstates and local density of states
(LDOS) in the energy representation. This novel approach has 
already been successfully applied to models of two interacting particles
\cite{Borgo,Benet,Lilia} and to a three orbital schematic shell
model \cite{Weng}. In this paper we extend this approach to chaotic
periodic billiards by incorporating the effects of the boundaries 
into the Hamiltonian operator and treating the degrees of freedom as
independent ``particles". Some results of this
work were advanced in \cite{Pla2000}.

The outline of the paper is as follows. Section II presents some
basic features of the classical dynamics of the rippled billiard,
together with the necessary details of its quantum description. 
In Section III the classical representation of the model is given.
In Section IV we discuss the meaning of the structure of
eigenstates (SES) and LDOS, and in Sections V and VI we define
their classical counterparts. 
In Section VII we compare the quantum and classical SES and LDOS. 
Section VIII pertains to the
individual properties of eigenstates and LDOS. In Section 
IX we make some clncluding remarks.

\section{Description of the model}

The system that we shall use to explore the quantum-classical
correspondence is a 2D billiard with periodic boundary conditions
along the longitudinal coordinate $x$. The top profile is given by
the function $y=d+ a \xi(x)$ with $\xi(x+2\pi)=\xi(x)$, where $a$
is the amplitude of the ripple and $d$ is the width of the
billiard when $a$ is zero. The bottom boundary is flat (see Fig. 1).

The first studies of the classical dynamics of this system 
were performed in Ref. \cite{Month} in
connection with the beam-beam interaction problem (see also \cite
{licht}). More recently, the finite length version of this system was
analyzed in Ref. \cite{crch} as a model of a mesoscopic electron
waveguide, where a transport signature of chaos in the behavior
of resistivity was established.
The same model has been also used \cite{Ketz} to study certain
{\it quantum} transport properties through ballistic cavities. 
On the other hand, the
analysis of the band-energy spectrum for an {\it infinitely} long
rippled channel \cite{qrch} in the cases of mixed and global
classical chaos, provides insight into the universal
features of electronic band structures of real crystals
\cite{Mucciolo}. In Ref.\cite{Dittrich} a similar periodic
billiard was considered in connection with the problem of chaotic
diffusion in the systems with band spectra, see also Refs.
\cite{lent,demi}.

An attractive feature of this rippled channel is that its
classical phase space undergoes the generic transition to global
chaos as the ripple amplitude $a$ increases. Hence results obtained for 
this particular system are expected to be applicable to a large class of 
systems, namely non-degenerate, non-integrable Hamiltonians.

In the quantum description the model is given by the Hamiltonian
\begin{equation}
\hat H =\frac{1}{2m_e}(\hat P_x^2 + \hat P_y^2)=
-\frac{\hbar^2}{2m_e}(\partial_x^2 +\partial_y^2)
\label{Ham0}
\end{equation}
for the wave function $\Psi(x,y)$ obeying the boundary conditions,
$\Psi(x,y)=0$, at $y=0$ and $d + a \xi(x)$. There exist various
numerical methods that can be used to obtain the eigenvalues and
eigenfunctions of non-integrable billiards, such as the transfer
matrix approach \cite{Sheng}, the scaling method
\cite{Vergini}, and the scattering approach \cite{Dittrich}. 
Here we shall employ a different technique that is
taylored to explore the quantum-classical correspondence of the
structure of eigenstates. It consists of representing the
Hamiltonian (\ref{Ham0}) in new coordinates ($u,v$) in such a way
that the effects of the boundary are transferred to an effective 
interaction potential between the two degrees of freedom $(u,v)$. That
is, the coordinates ($u,v$) are chosen so that the new wave
function satisfies ``flat" boundary conditions: $\Psi(u,v)=0$ at
$v=0$; $d$. For our rippled channel, this can be accomplished by the
transformation
\begin{eqnarray}
u & = & x \nonumber \\ v & = & \frac{y d}{d+ a \xi(x)}=
\frac{y}{1+\epsilon \xi(x)},
\label{trans}
\end{eqnarray}
where $\epsilon \equiv a/d$ is a measure of the perturbation due
to the ripple \cite{note2}. The Schr\"odinger equation in the new
coordinates can be obtained from the covariant expression for a
particle moving (in the absence of potentials) in a Riemannian
curved space \cite{Adler},
\begin{equation}
-\frac{\hbar^2}{2m_e} \Delta_{cov} \Psi (u,v) =
\frac{\hbar^2}{2m_e} g^{-1/2}
\partial_\alpha g^{\alpha \beta}
g^{1/2} \partial_\beta \Psi (u,v).
\label{Delta}
\end{equation}
Here $\Delta_{cov}$ is the Laplace-Beltrami operator, $g$ is the
metric and $g^{\alpha \beta}$ is the metric tensor. Even though
(\ref{Delta}) is still the kinetic energy, the resulting differential
equation takes a much more complicated form than the ordinary
Laplacian. This is the price we have to pay when we transfer the
effect of the boundaries onto the operator [the explicit form of the
Schr\"odinger equation in $(u,v)$ coordinates for the rippled
billiard is given in \cite{qrch}]. Moreover, the coordinate
representation of the canonical momentum has now the form
\cite{Dewitt},
\begin{equation}
\hat P_\alpha=-i \hbar\left[\partial_\alpha + \frac{1}{4}
\partial_\alpha ln(g)\right]=
-i\hbar g^{-1/4}\partial_\alpha g^{1/4}.
\end{equation}

Substitution of this expression into the Schr\"odinger equation
(\ref{Delta}) determines the quantum Hamiltonian in covariant form
\cite{Dewitt}
\begin{equation}
\hat H= \frac{1}{2m_e} g^{-1/4}\hat P_\alpha g^{\alpha \beta} g^{1/2}
\hat P_\beta g^{-1/4}, \ \ \alpha, \beta=u,v,
\end{equation}
Substituting now the explicit expressions for the metric tensor
$g^{\alpha \beta}$ and metric $g$,
\begin{equation}
g^{\alpha \beta} = \left( \begin{array}{cc}
1 & \frac{-\epsilon v \xi_u}{1+\epsilon \xi} \nonumber \\
\frac{-\epsilon v \xi_u}{1+\epsilon \xi} & 
\frac{1+\epsilon^2 v^2 \xi_u^2}{(1+\epsilon \xi)^2} 
\end{array} \right)
\label{tensor}
\end{equation}
\begin{equation}
g = Det(g_{\alpha \beta}) = [1 + \epsilon \xi u]^2,
\end{equation}
we get
\begin{eqnarray} 
\hat H & = & \frac{1}{2m_e} g^{-1/4} \left\{\hat P_u (1+\epsilon \xi) \hat P_u +
\hat P_v \frac{1 + \epsilon^2 v^2 \xi_u^2}{1 + \epsilon \xi} \hat P_v
\right. \nonumber\\
& - & \left. \epsilon \left[ \hat P_u v \xi_u \hat P_v + \hat P_v v \xi_u \hat P_u 
\right] \right\} g^{1/4}
\label{Ham1}
\end{eqnarray}
where $\xi_u=d\xi/du$. Note that the rippled
boundary is fully incorporated into the Hamiltonian operator.
Since the boundary conditions in the new coordinates correspond
to those of a ``flat" channel, $\Psi(u,v)=0$ at $v=0$; $d$. For our
purposes it is convenient now to separate the Hamiltonian (8) as
\begin{equation}
\hat H = \hat H^0 + \hat V(u,v, \hat P_u, \hat P_v),
\end{equation}
where
\begin{equation}
\hat H^0=\frac{1}{2m_e}(\hat P_u^2 + \hat P_v^2)
\label{H00}
\end{equation}
with [c. f. (4) and (7)]
\begin{equation}
\hat P_u = -i \hbar \left[ \partial_u + \frac{1}{4}
\partial_u ln(g) \right],\
\ \hat P_v = -i \hbar \partial_v,
\end{equation}
and $\hat V$ stands for the rest of the terms.

We remark that this representation allows us to treat the original
model of one free particle in the rippled channel as a 1D model of
two interacting ``particles" identified with the two degrees of
freedom $u$ and $v$. Here $H^0$ is the Hamiltonian of two {\it
non-interacting particles}. The eigenfunctions of $H^0$ define the
unperturbed basis in which the eigenstates of the total
Hamiltonian $\hat H(u,v,\hat P_u, \hat P_v)$ may be expanded. Such a
representation turns out to be convenient for the study of chaotic
properties of the model since one can use the tools and concepts
developed in the theory of interacting particles (see, for
example, \cite{FI97}).

Since the Hamiltonian (\ref{Ham1}) is periodic in $u$, the
eigenstates are Bloch states. This allows us to write the solution of
the Schr\"odinger equation in the form $\Psi_E(u,v)=\exp
(iku)\Phi_k(u,v)$ with $\Phi_k(u+2\pi,v)=\Phi_k(u,v)$. For an
infinite periodic channel the Bloch wave vector $k=k(E)$ takes a
continuous range of values, which we take to lie on the first
Brillouin zone ($-\frac{1}{2}\leq k \leq \frac{1}{2})$.

By expanding $ \Phi_k(u,v)$  in a double Fourier series the
$\alpha$-th eigenstate of energy $E^{\alpha}(k)$ can be written as
\begin{equation}
\Phi^{\alpha}(u,v;k)= \sum_{m=1}^{\infty}
\sum_{n=-\infty}^{\infty}C_{mn}^{\alpha}(k) \phi_{mn}^k(u,v),
\end{equation}
where
\begin{equation}
\begin{array}{ll}
\phi_{mn}^k(u,v)\,=\,<u,v \mid m,n>_k=
\\ ~~~~~~~~~\\
\pi^{-1/2}g^{-1/4}\,\,\sin(m \pi v/d)\,\,
\exp\left[{i(k+n)u}\right]
\end{array}
\label{phi}
\end{equation}
are the eigenstates of the unperturbed Hamiltonian $\hat H^0(u,v)$
with corresponding eigenvalues
\begin{equation}
E^{(0)}_{n,m}(k) =
\frac{\hbar^2}{2m_e}\left[(k+n)^2+\left(\frac{m\pi}{d}\right)^2\right].
\label{E00}
\end{equation}
The factor $\pi^{-1/2} g^{-1/4}$ in Eq. (\ref{phi}) appears from
the orthonormality condition in the curvilinear coordinates
$(u,v)$. Thus, the exact eigenstates are expanded in a complete
orthonormal basis satisfying the boundary conditions of the
problem, {\it i.e.}, a Galerkin series expansion \cite{RLI}.

So far, there has been no need to specify the ripple function
$\xi(x)$, except that it is periodic: $\xi(x+2\pi)=\xi(x)$. 
For concreteness, we consider from now on the dependence
$\xi(x)=\cos(x)$. The ripple profile is given by $y=d + a
\cos(x)$ where $x$, $y$, $d$, and $a$ are dimensionless
quantities. The latter are defined from the expression $Y=D + A
\cos(2\pi X/B)$ scaled to the period $B$. Therefore, $x= 2\pi \frac{X}{B}$,
$y=2\pi\frac{Y}{B}$, $d= 2\pi \frac{D}{B}$, and $a= 2\pi
\frac{A}{B}$, where $X$, $Y$, $D$, $A$, and $B$ are dimensional
quantities.

This approach to solving the Schr\"odinger equation was used to
calculate the energy-band structure and Husimi distributions of
the rippled channel \cite{qrch}, and to obtain the energy level
statistics under the conditions of full and mixed chaos
\cite{rmf}. Specifically, in Ref.\cite{rmf}
the energy level spacing distribution was shown to be Poisson,
Wigner-Dyson, or intermediate between these two, for regular,
globally chaotic, or mixed classical motion, respectively, in 
agreement with the well known RMT conjecture \cite{conjecture}. Most
relevant to our work here is the fact that these distributions
were found to be the same for all values of $k$ within the
Brillouin zone, except $ k \approx 0$ for which the parity
symmetry should be taken into account. Thus, without loss of
generality we fix the value  $k=0.1$. We
shall also use in all our calculations the following values for
the ampliutude $a$ and width $d$ of the rippled channel: $a/2
\pi=0.06$ and $d/2 \pi = 1.0$. These values produce
representative global chaotic motion in the classical limit (see
below).

In order to study the structure of eigenstates of the total
Hamiltonian $\hat H(u,v)$ one needs, first, to choose a way of
ordering the unperturbed basis in which to represent
the Hamiltonian matrix
$H_{l,l'}(k)= <l\mid\hat H\mid l'>_k$.
Specifically, we have to assign an index $l$, labeling the basis
state $\mid l>_k \equiv\mid m,n>_k$, to each pair of indecies
$(m,n)$ (note that, although the energy spectra is
independent of the assignment $(m,n)\rightarrow l$, the structure
of the eigenstates is not). The size of the Hamiltonian 
matrix is determined by the
maximum values of $n$ and $m$: $-N_{max}\leq n\leq N_{max}$ and $1
\leq m \leq M_{max}$.

A natural assignment is the following one. Let us fix the lowest
value of $n$ $(-N_{max})$ and sweep all values of $m \quad (1 \leq m \leq
M_{max})$. This gives $l=1, 2, ..., M_{max}$. Then do the same for
$n=-N_{max}+1$, which gives $M_{max}+1 \leq l \leq 2M_{max}$, and
so on, till finally we have $1 \leq l \leq L_{max}$, where
$L_{max}=(2N_{max}+1)M_{max}$ defines the matrix size, $L_{max}
\times L_{max}$. This rule results in a block structure of the
Hamiltonian matrix, with block size equal to $M_{max}$. Fig. 2
shows the central part of a $4030 \times 4030$ matrix with
$(N_{max},M_{max})=(32,62)$. Here we can see a number of blocks of
size $62\times 62$ corresponding to $n,n'=0,\pm 1, \pm 2
\pm 3$. In this representation the matrix is clearly a band matrix.
The finite size of the band is due to the short range coupling
between different blocks, namely, the strength of
the matrix elements decrease as
$\epsilon ^{\mid n-n'\mid}$, $\epsilon ^{\mid n - n'\pm1\mid}$, or
$\epsilon ^{\mid n-n'\pm 2\mid}$ (see details in \cite {qrch}).

The above way of ordering the unperturbed basis is typical; it
corresponds to the ``channel representation" since the index $m$
labels a specific transverse channel (or mode) for the propagation of the
wave through the billiard, see Eq. (\ref{phi}). However, for our
purposes it is essential to use the ``energy representation",
according to which the unperturbed basis is ordered in increasing
energy, $E^0_{l+1}(k)\geq E^0_l(k)$. This defines a new rule
$l \rightarrow l_{new}=l_{new}(n,m)$.  Fig. 3 shows the difference
between these two ways of ordering of the basis, giving the
normalized unperturbed energy $\bar{E}^0_l(k)\equiv
E^0_l(k)/E^*$=$(\frac{d}{\pi})^2(n+k)^2 + m^2$ with $E^* \equiv
\frac{\hbar^2}{2}\frac{\pi^2}{d^2}$, as a function of $l$.

A non-monotonic dependence of the energy occurs in Fig. 3a because
the first rule $l(m,n)$ described above produces
minimum values in the energy for $m=1$. In contrast, in Fig. 3b a
linear dependence of the unperturbed energy on the index $l_{new}$
is shown, apart from large values of $l_{new}$ where the
finiteness of the matrix results in the distortion of the spectra.
The numerical obtained linear dependence, $\bar{E}\equiv E/E^*=\frac{4}{\pi}
l_{new}$, can be derived analytically from Weyl's formula
$\bar{N}(E)=\frac{S E m_e}{2 \pi\hbar^2}$, valid for 2D billiards with
area $S$ (also valid for {\it periodic} 2D billiards)
\cite{Note3}.

A crucial point is that the eigenstates of the total Hamiltonian
in the ``energy representation" have a very convenient form for
the analysis. The advantage of the ``energy representation" is
clearly seen in Fig. 4 where an arbitrarily chosen eigenstate is
given in the two representations. One can see that in one case the
eigenstate has a kind of regular and extended structure, while in
the other, the eigenstate is compresed around the unperturbed
state whose energy is close to the energy of the perturbed state. In the
latter case one may use a statistical approach to describe the
global properties of such eigenstates, see
Refs.\cite{FI97,FI00,I00}. Specifically, we may characterize such
eigenstate by introducing an envelope around which the components
are expected to fluctuate in a pseudo-random way. We stress that 
by using this
energy ordering one can relate the global form of eigenstates in
the energy representation with its classical counterpart, see
below.

\section{Classical representation of the model}

Clearly, the type of motion of a particle in the ripple billiard
depends on the values of the geometrical parameters, namely, on
the ripple amplitude $a$ and the average width $d$. If the channel
is narrow, $ d \ll 2\pi$, the Poincare section reveals a large
resonance island surrounded by the typical Poincare-Birkhoff
structure \cite{Month,crch,qrch}. This island is formed by the
librational motion along the $x$ direction in the neighborhood of
$x=0$. On the contrary, for wide channels ($d \geq 2\pi$) global
chaos occurs even for small amplitudes due to a strong overlap of
resonances. In this work we shall limit ourselves to the study of
global chaos, {\it i.e.}, wide channels.

The condition of global chaos can be derived analytically for the
case of small amplitude $a \ll d$ where the following approximate
map is valid \cite{Note1},
\begin{equation}
\begin{array}{rcl}
\alpha_{n+1} & = & \alpha_n + 2a\, \sin(x_n)\\
x_{n+1} & = & x_n + 2 d \, \cot(\alpha_{n+1}),\quad (mod
\,\,2\pi),
\end{array}
\end{equation}
Here $x_n$ is the position of a particle and $\alpha_n$ is the
angle between the $x$-axis and the velocity of the particle, right
after the $n-$th collision with the upper wall. The standard
linearization around fixed points of period one yields,
\begin{equation}
\begin{array}{ccl}
\delta \alpha_{n+1} & = & \delta \alpha_n + \kappa \, \sin(\Delta x_n)\\
\Delta x_{n+1} & = & \Delta x_n + \delta \alpha_{n+1},\quad (mod \,\,2\pi),
\quad \kappa \equiv 2 d a/\pi
\label{map1}
\end{array}
\end{equation}
where the angle $\delta \alpha$ and the position $\Delta x$,
measured from the fixed point, play the role of the action and
angle variables, respectively. Eq. (\ref{map1}) has the form of the
{\it standard map} with $\kappa$ as the nonlinear parameter.
Hence, Chirikov's overlapping criteria predicts the onset of
global chaos for $\kappa \simeq 1$. This was confirmed by
computing numerically the actual path of the particle as it
travels along the channel. Specifically, for $\kappa \geq 1$ the
Poincare section $(\alpha_n, x_n)$ shows global chaos for wide
channels. In this paper we shall consider a wide channel with 
small ripple amplitude, specified by the parameters
$(a/2\pi, d/2\pi)=(0.06,1.0)$, therefore, $\kappa \approx 1.5$.

Let us now write, as in the quantum description, the classical
Hamiltonian in the $(u,v)$ coordinates so that the effects of the
rippled boundary can be incorporated as a kind of coupling
between two degrees of freedom. The classical Hamiltonian is
\begin{equation}
H= \frac{1}{2m_e} g^{\alpha \beta} P_{\alpha}P_ {\beta},
\label{Ham2}
\end{equation}
which can be obtained from the quantum Hamiltonian (\ref{Ham1}) by
commuting all momenta and coordinates \cite{NOTE}.

For completeness, the canonical transformation between
$(x,y,P_x,P_y)$ and $(u,v,P_u,P_v)$ are given by Eq. (\ref{trans})
together with $P_u = P_x + yP_y/(1+\epsilon \xi)$, and
$P_v=(1+\epsilon
\xi)P_y$.
Inserting the metric tensor Eq. (\ref{tensor}) into (\ref{Ham2}) and
regrouping terms, the Hamiltonian can be written as $H = H^0 + V$,
where
\begin{equation}
H^0 = \frac{1}{2m_e}(P_u^2 + P_v^2)
\label{H0cl}
\end{equation}
and
\begin{equation}
V=-\frac{1}{2m_e}\epsilon \left[\frac{2 \xi + \epsilon \, (\xi^2 +
v^2
\xi_u^2)}{(1+\epsilon \xi)^2}P_v^2 +
\frac{2 \xi _u}{1+\epsilon \xi} P_u P_v \right].
\label{Vcl}
\end{equation}
V represents the interaction between the two "particles" $u$ and $v$ 
(here $\xi_u = d\xi/du$, and $\epsilon =a/d$).

\section {Structure of eigenstates and LDOS: definitions.}

Once the matrix $H_{l,l'}(k)$ has been diagonalized, its
eigenstates $\Psi^{\alpha}(k)=\sum C_l^{\alpha}(k) \phi^k_l$ are
also re-ordered in energy: $E^{\alpha +1}\geq E^{\alpha}$. We
adopt  the convention that the greek superindex (latin subindex)
denotes the exact (unperturbed) state. The amplitudes
$C^{\alpha}_l(k)$ form the ``state vector matrix". The elements
along the row $\alpha$ of this matrix are the components of the
$\alpha^{th}$ eigenstate in the representation of the unperturbed
energy-ordered basis. Correspondingly, the elements along the
column $l$ give the unperturbed state $l$ expanded in the energy 
ordered perturbed basis. For our purposes, we examine the
matrix $w_l^\alpha \equiv |C^{\alpha}_l(k)|^2$ (see Fig. 5) which
plays the central role in our approach.

The rows of the matrix $w_l^\alpha$ show how a specific
eigenfunction (eigenstate $|\alpha>$) is expanded in the
unperturbed basis $|l>$. Since the average ``band width'' of
$w^\alpha_l$ smoothly depends on the index $\alpha$ (equivalently
on the energy $E^\alpha$), it is convenient to average
$w_l^\alpha$ over a small range of eigenstates. In this way, one
can obtain the so-called {\it structure of eigenstates} (SES) in the
unperturbed basis. In what follows, we are interested in the SES
in the energy representation which is defined as \cite{CCGI96},
\begin{equation}
W(E_l^0 \mid E^\alpha)=
\sum_{\alpha'} \bar{w}_l^{\alpha'} \,\delta(E_l^0-E^{\alpha'}),
\label{sef}
\end{equation}
where $\bar{w}_l^{\alpha'}=w_l^{\alpha'}/N$ with $N$ as the number
of eigenstates $|\alpha'>$ in the vicinity of a given $\alpha$. 
The structure of eigenstates $W(E_l^0 \mid E^\alpha)$
gives the dependence on the unperturbed energy $E_l^0$ for 
eigenstates with total energy close to $E^\alpha$. This sum can be
approximated by an integral that involves the density of states
$\rho(E^\alpha)$ providing the transition from the basis to the
energy representation. This function SES plays the key role in the
determination of important physical quantities, such as the
distribution of occupation numbers in chaotic closed systems (see,
for example \cite{Benet,FI97,BGIC98}).

Analogously, one can analyze the columns of the matrix
$w_l^\alpha$ which give information about a given {\it
unperturbed} state $|l>$ is spanned over {\it exact} eigenstates
$|\alpha>$. Therefore, one can define the following quantity
\begin{equation}
\omega(E^{\alpha} \mid E^0_l) = \sum_{l'} \bar{w}_{l'}^\alpha \,
\delta (E^{\alpha}-E^0_{l'}),
\label{ldos}
\end{equation}
which is well known in nuclear physics as the {\it strength
function}, and in solid state physics, as the {\it local density
of states} (LDOS). In (\ref{ldos}) the summation (average) is done
over $N$ {\it basis} states $|l'>$ around a fixed state $|l>$,
$\bar{w}_{l'}^\alpha=w_{l'}^\alpha/N$. Therefore, this quantity is
considered as a function of the energy $E^\alpha$. It shows
how a given unperturbed state $\mid \phi_l>$ is coupled to the
exact states due to the interaction $V$. The width of this
function ({\it spreading width}) determines the energy range
associated with the ``decay" of a given unperturbed state into
other states when the interaction is switched on.

\section {Classical analog of the structure of eigenfunctions}

An essential point is that both, the structure of eigenstates and the
local density of states, have well defined classical analogs
\cite{CCGI96}. Let us first start with the SES. Since
$C^{\alpha}_l =<\Psi^{\alpha}\mid\phi_l>$ as a function of $l$ is the
{\it projection} of the perturbed state onto the states of the
unperturbed system, the classical counterpart of $w_l^\alpha =
\mid C^{\alpha}_l\mid^2$ as a function of energy $E_l$ can be
defined  as the {\it projection} of the total Hamiltonian $H$ onto
the unperturbed one $H^0$, where $H=H^0 +V$ with $V$ the
perturbation \cite{CCGI96}. This projection can be numerically
done by substituting the trajectories $\Phi(t)\equiv (x(t), y(t),
p_x(t), p_y(t))$ generated by the Hamiltonian $H$ with energy $E$
onto the unperturbed Hamiltonian $H^0$. Since the unperturebed
energy $E^0(t)$ along these trajectories varies in time, it fills
the so-called {\it energy shell} characterized by its width
$\Delta E$. Therefore, for chaotic total Hamiltonians $H$, the
classical analog $W_{cl}(E^0 \mid E)$ of the quantum SES can be
easily obtained from $E^0(t)$.

In Fig. 6a we show the energy of the unperturbed Hamiltonian $H^0$
as a function of time after the substitution of a single chaotic
trajectory $\phi(t)$ generated by the perturbed Hamiltonian $H$,
into $H^0$. The classical distribution $W_{cl}(E^0
\mid E)$ (see Fig. 6b) is constructed from $E^0(t)$, averaged over
a sufficiently long time. Alternatively, since the dynamics of $H$
is chaotic, the same distribution can be obtained by averaging
over many different orbits at a shorter time. For concreteness, in
what follows we chose $E=1$.

It is important to understand the origin of the shape of the
classical distribution, see Fig. 6b, in view of its correspondence
to the quantum SES. For this, let us examine in detail
characteristic time intervals of the plot of $E^0(t)$ together
with plots of $P_u^2$, $P_v^2$, and $v$, see Figs. 7a-d. From
Fig. 6a it is clear that the main contribution to the central peak
of the distribution $W_{cl}(E^0 \mid E)$ comes from the continuous
time intervals where $E^0 \approx 1$. These intervals in turn
correspond to $P_u^2 \approx 2$ (the maximum value) and to $P_v^2
\approx 0$ (see Figs. 7b-c). During these time intervals the particle 
travels almost
parallel to the $x$ axis ($u$ axis). In fact, from the plot of
$v(t)$ (Fig. 7d) one can see that the particle hits only once the
bottom boundary during a relatively long time. The contribution to
the central peak can be also be seen directly from the expressions
(\ref{H0cl}) and (\ref{Vcl}). Indeed, since $P_v^2 \approx 0$, we
have  $E^0(t) \approx \frac{1}{2} P_u^2 \approx E = 1$.

In the same way one can understand the origin of two side-band
peaks in the form of $W_{cl}(E^0 \mid E)$, clearly seen in Fig. 6b.
The data reported in Fig. 8a-d show that for small time intervals
e. g., ($0.212 < t < 0.216$ and $0.231 < t < 0.234$) the value of $E^0$
is practically constant ($E^0 \approx 1.12$ and $E^0 \approx 0.88$
resepectively). These values of $E^0$ correspond to the right and
left peaks in Fig. 6b. Contrary to the data in Fig. 6 explaining the
origin of the central peak, in the two time regions of Fig. 8 the
value of $P_u^2$ is nearly zero and $P_v^2$ is nearly maximum (see
Fig. 8b-c). In these regions the motion of the particle is almost
perpendicular to the $x$-axis ($u$-axis).

The form of the profile $\xi(u) = \cos(u)$ in our ripple channel
results in two periodic orbits of period one: the stable one for
$P_u=0, u=0$, and the unstable one for $P_u=0, u=\pi$. Putting
these values for $P_u$ and $u$ into expressions (18-19), we find
the values for $P_v^2$. This allows us to obtain the value of $H^0$
for these two periodic orbit, $E^0=E(1+\epsilon)^2 \approx 1.12$
and $E^0=E(1-\epsilon)^2\approx 0.88$, respectively for $u=0$ and
$u=\pi$. Thus, the left (right) peak of the distribution
$W_{cl}(E^0 \mid E)$ is formed by trajectories dwelling near the
unstable (stable) periodic orbits of period one.

\section {Classical analog of the LDOS}

In analogy with the quantum LDOS, the classical LDOS distribution
$\omega_{cl}(E \mid E^0)$ is constructed by projecting the
dynamics generated by $H^0=E^0$ onto the Hamiltonian $H=H^0 +V$.
Here, because $H^0$ is integrable, several ({\it regular})
trajectories of $H^0$ have to be substituted into $H$ and then
averaged in time, in contrast with the classical SES where only
one ({\it chaotic}) trajectory over a sufficiently long time is
needed to form the distribution. The result of this procedure is
exemplified in Fig. 9a where $E(t)$ is shown for $36$ regular
trajectories, whose initial conditions were chosen to lie on a
mesh of points distributed uniformily in the plane ($P_u,u$) ($v$
is fixed and $P_v$ is determined by the constant energy $E^0=1$).
The classical LDOS distribution $\omega_{cl}(E \mid E^0)$, see
Fig. 9b, was obtained from $E(t)$ (Fig. 9a).

In the construction of Fig. 9a, $6$ initial conditions for $u$
(taken from the interal [$-\pi$,$\pi$]) were used for each of the
$6$ initial values of $P_u$ [from the interval
($-\sqrt{2E^0}$,$\sqrt{2E^0}$)]. Hence the appearance of $6$
similar sets formed by $6$ small time intervals. The very first
time interval ($[0,0.25]$) yields $E=1$ and is produced by a
trajectories with initial condition $P_u \stackrel{>}{\sim} 
-\sqrt{2}$ and $u=-\pi$, {\it i.e.}, a grazing trajectory.
The next interval $[0.25,0.5]$, a bar code-like structure
extending over the whole energy shell, is produced by the initial
condition $P_u=-0.6\sqrt{2}$, $u=-\pi$, and the third interval
$[0.5,0.75]$, the noise-like one, is produced by the initial
condition $P_u=-0.2\sqrt{2}$, $u=-\pi$. The next three intervals
are produced by mirror initial conditions ($P_u=0.2\sqrt{2}$,
$0.6\sqrt{2}$ , and $\stackrel{<}{\sim} \sqrt{2}$ with $u=-\pi$). 
The initial
conditions for the next sets are the same as for the first set,
except $u$ is shifted subsequently by $2\pi/5$. Any trajectory
reveals one of these three types of behavior of $E(t)$; $E=1$, bar
code, and noise-like structure. The averaging over these time
intervals yields the distribution shown in Fig. 9b, quite different
from the classical SES (Fig. 6) although it also has the central
peak and two other peaks on each side.

\section{Quantum-classical correspondence: global properties}

Having understood the dynamical origin of the main features of the
classical distributions $W_{cl}(E^0 \mid E)$ (SES) and
$\omega_{cl}(E \mid E^0)$ (LDOS), we now compare them with the
corresponding quantum quantities. Although the SES for a given
$\alpha$, see Eq. (20), is actually the average over a set of
perturbed states in the neighborhood of $\alpha$, it is
convenient, first, to examine typical {\it indvidual} states in
different energy regions. In order to compare the
quantum and classical behavior, we use the quantum energy of
interest $E_q$ (where $E_q$ is $E^\alpha$ in the case of SES
and $E^0_l$ in the case of LDOS) to construct
the classical distributions ($E_{cl}=E_q$). The ratio of the de 
Broglie wavelength $\Lambda =2\pi \sqrt{\frac{\hbar^2}{2m_eE}} $ to some
characteristic length of the billiard may be used as a semiclassical 
parameter (recall that there are two length parameters defining the
flat channel, $D$ and $B$; in this work we have used $D=B$). 
For example $\Lambda/D=2\pi \hbar_{eff}<<1$, where 
$\hbar_{eff}\equiv \frac{\hbar}{D\sqrt{2m_eE}}$, tells us how
large the energy of the eigenstate should be for its wavelength to 
be  many times smaller than the width of the channel. Substitution 
of Weyl's formula $\bar N(E)=\frac{BD  m_e}{2\pi \hbar^2}E$ into
the above expression for $\Lambda$ gives a more useful formula:
\begin{equation}
\Lambda/D= \sqrt{\frac{B}{D}\frac{\pi}{\bar N}}= 2\pi \hbar_{eff}.
\end{equation}
Here we only need to know the number of the eigenstate (as oppossed 
to the energy) to determine the ratio $\Lambda/D$.

Fig. 10 shows $w^{\alpha}_l$ for four representative states ($\alpha = 
111, 712, 1362, and 2002$) of different energy regions as a function 
of the unperturbed energy $E^0_l$ scaled to $E^\alpha$. The classical 
SES is also shown. As it may be expected, these figures show that as 
$\hbar_{eff}$ becomes smaller the number of unperturbed states needed 
to form the pertubed state increases. Specifically, level states around 
$\alpha=111$ are still far from the semiclassical regime 
($\Lambda/D \approx 0.17 $); the details of the rippled boundary 
cannot be resolved and, thus, there is only a weak mixing of the 
unpertubed states. In contrast, for the state $\alpha=2002$ its de 
Broglie wavelength is a smaller fraction of the width of the channel 
($\Lambda/D\approx 0.04$) and the number of participating components 
is larger.

It is important to note that although the number of unperturbed
states needed to construct the perturbed state increases as the
energy increases, they all fall within the range of energies
determined by the classical distribution. This range defines the
{\it energy shell} of the eigenstate under consideration. On the 
other hand, it is clear that even the level $\alpha=2002$ does 
not seem to have much of a resemblance with the classical 
distribution. For this and similar states, we have calculated
the distribution of fluctuations arround the classical curve
and found that they deviate strongly from a Gaussian distribution
as predicted by RMT considerations. This may be related to the 
fact that the de Broglie wavelength of level 
2002 is still not much smaller than the amplitude $A$ of the ripple 
($\Lambda/A\approx 0.65$, recall A/D=0.06). That is, even though 
$\Lambda/D$ may be small, a more relevant parameter to
characterize the semiclassical regime for particular eigenstates 
is $\Lambda/A$. Nevertheless the strong fluctuations exemplified by the 
level $\alpha=2002$ (Fig. 10d) can be smoothed out by averaging over a 
range of perturbed states. This actually corresponds to the definition 
(\ref{sef}) of the structure of eigenstates $W(E^0_l \mid E^\alpha)$.

Remarkably, as Fig. 11 shows, such an average does reveal a good
quantum-classical correspondence even though the fluctuations are
strong. Specifically, the global shape
of the quantum SES displays a three-peak structure, much like the
classical one. In addition, the energy spread of the quantum SES
agrees very well with the classically determined energy shell,
even in its slight assymetry ($ 0.7<E^0<1.45$ with the center at
$E^0=1$).

Similarly, in Fig. 12 we compare the quantum and classical LDOS.
The quantum-classical correspondence appears to be even better
than for the SES in the sense that the fluctuations are smaller.

\section{Localization and non-ergodicity}

To understand the origin and importance of the strong fluctuations
of the SES around the classical counterpart (see Fig. 11), we
analyze the structure of {\it individual} eigenfunctions. In
Fig. 13 three consecutive typical eigenfunctions ($\alpha=1952$,
$1953$, and $1954$) are shown. The difference between the state
$\alpha=1954$ and the other two is clearly qualitative. More 
specifically, while the states $\alpha=1952$ and $1953$ are {\it 
extended} (in energy) states, constituted by practically all 
unperturbed eigenstates within the energy shell, the state 
$\alpha=1954$ is mostly unperturbed; it is extremely {\it localized} 
in the energy shell. By neglecting all small amplitude components 
surrounding the main component (see Fig. 13c), we can determine 
the unperturbed state, defined by a
pair $(m,n)$, that most closely resembles the perturbed state. We
find that this always corresponds to the lowest values of the
transversal mode $m$. This fact can be understood by the following 
physical argument. Consider an eigenstate of the flat channel 
$\phi^0_{m,n}(X,Y)\propto \sin(m\pi Y/D)\exp(iKX)$ with energy 
$E^0= \frac{\hbar^2}{2m_e}(K_x^2+ K_y^2)$, where 
$K_y=\frac{m\pi}{D}=\frac{2\pi}{\Lambda_y}$. Turning on the 
perturbation (flat to rippled channel) will affect the high energy 
unperturbed states differently depending mainly on the value of 
$\Lambda_y$. For example, for $m=1$, the ratio $\Lambda_y/A$ is 
$2D/A \approx 33$, which is so large that the state cannot "see" 
the ripple and thus will remain essentially unperturbed. In contrast, 
for unperturbed states with the same (or about the same) energy but 
with large values of $m$ (say, $m=62= M_{max}$ and correspondingly 
small $k_x$) their $\Lambda_y$ is sufficiently small, compared to the
amplitude of the ripple ($\Lambda_y/A\approx 0.5$), that the rippled 
boundary produces a strong mixing of unperturbed levels. The 
resulting perturbed state will consist of many components, extended 
over the energy shell. By the same token, it is expected
that unperturbed states with intermediate values of $m$ will turn, 
after turning on the perturbation, into some intermediate states, 
the so-called sparse states [32], made up of fewer components over 
the energy shell.

The existence of the extremely localized (in energy) states manifests 
itself in the structure of the Hamiltonian matrix $H_{l,l'}$ in the 
channel representation (Fig. 2) to be discussed below. It is easy 
to find that such wavefunctions differ only slightly from the plane 
waves $\phi^0_{m,n}(x,y)$ with small $m$, proper of the flat 
channel. In Fig. 14a one localized (in energy) state is shown in 
configuration representation. In contrast, an extended (in energy) 
state is presented in Fig. 14b.

Eigenfunctions like that of Fig. 13b for $\alpha=1953$, are
somewhat intermediate between the extended and localized states.

In order to characterize quantitatively the eigenfunctions, we compute
various localization measures. The first one is the so-called {\it
entropy localization length} $l_H$,
\begin{equation}
l_H =  \exp {\left[-\left( \cal H - \cal H_{GOE}\right)
\right]} \approx 2.08 \exp {(-\cal H)}.
\label{shennon}
\end{equation}
Here $\cal H$ stands for the Shannon entropy of an
eigenstate in a given basis,
\begin{equation}
{\cal H}=\sum^N_{l=1} w_l^\alpha \,\ln w_l^\alpha,
\label{shannon}
\end{equation}
and $\cal H_{GOE}$ is the entropy of a completely chaotic state
which is characterized by gaussian fluctuations (for $N
\rightarrow \infty $) of all components $C_l^\alpha$ with the same
variance $<w_l^\alpha>=1/N$. The latter property occurs for
completely random matrices belonging to a Gaussian Orthogonal
Ensemble (GOE). Defined in this way, the quantity $l_H$ gives a
measure of the effective number of components in an eigenstate. 
For example, the eigenstates of Fig. 13 have, respectively, 
$l_{\cal H}=745.9$, $232.9$, and $2.7$.

The second quantity, $l_{ipr}$, which gives another
measure of the effective number of components in an eigenstate, is
expressed via the {\it inverse participation ratio} $\cal P$,
\begin{equation}
l_{ipr} = \left[ \frac {\cal P_{GOE}}{\cal P} \right]
\approx \frac {3}{\cal P},
\label{lipr}
\end{equation}
with
\begin{equation}
{\cal P}= \sum^N_{l=1} (w_l^\alpha)^2.
\label{ipr}
\end{equation}
where $\cal P_{GOE} \approx$ 3 is chosen in order to get $l_{ipr}=N$ 
in the GOE limit case. 

The above two definitions of ``localization lengths" are the most
frequently used when describing global structure of
eigenfunctions. One should note that these quantities give an
effective number of large components, independently on the
location of these components in the unperturbed basis.

Additional information about the structure of eigenfunctions can
be obtained from the ``width"  or mean square root $l_{\sigma}$ of an 
eigenstate, computed as
\begin{equation}
l_\sigma= \left[ \sum^N_{l=1} w_l^\alpha
\,\left[l-n_c(\alpha)\right]^2\right]^{1/2}
\label{sigma}
\end{equation}
where $n_c = \sum_l l \, w_l^\alpha$ determines the centroid of an
eigenstate in the unperturbed basis.

To get a complete panorama, in Fig. 15a-c we plot these three
measures as a function of $\alpha$ corresponding to the eigenstate
$|\alpha>$. The strong fluctuations of all localization measures 
are evident in these figures. We can see that neighboring high 
levels may have drastically different localization measures,
in agreement with the discussion above about the existence of the
three types of states: localized, extended, and sparse. These figures 
gives us information about the relative number of each type to be 
found in a given energy range.

Comparison of the width $l_{\sigma}$ with $l_H$ and $l_{ipr}$
gives the possibility to detect the so-called {\it sparsity} of
eigenstates. Indeed, small values of the ratio $l_H /l_{\sigma}$
(or $l_{ipr}/l_{\sigma}$) indicates that there are many ``holes"
in the structure of eigenstates, therefore, such eigenstates are
{\it sparse} \cite{robnik}. A detailed analysis shows a 
dominance of sparsed eigenstates. As for the centroid
$n_c=\alpha$ small fluctuations (observed in Fig. 15c) indicate that the
interaction strength $V$ is relatively weak, compared to the
unperturbed part $H_0$.

The data of Fig. 15 show the existence of some pattern for all
localization measures, visible as clusters of points in the lower
part of the plots. These patterns are much more pronounced for
the same quantities computed for the LDOS, see Fig. 16. In order to
understand this unexpected phenomenon, let us take a closer look at
$l_{\sigma}$ in the range $1450 < l < 2100$ and examine the
structure shown in Fig. 17 in detail. Specifically, we choose some
eigenstates from three branches of points (see marked states in 
Fig. 17). From each of these three sets,
we take the $1^{st}$, $7^{th}$, $13^{th}$, and $19^{th}$ states,
counted from the bottom of the figure, and plot them in Fig. 18.
Inspection of Figures 18a-18d clearly demonstrates that there is a 
kind of regularity in the structure of the LDOS: the same type of 
states appear repeatedly, almost periodically as a function of the 
basis number $l$. These figures show the repetition of extremely 
localized states (Fig. 18a) and of states with different degrees 
of sparsity (Figs. 18b-18d). The repetition of extended states, 
corresponding to larger vaues of $l_\sigma$ are not shown for 
economy of space but their existence is clear, as can be inferred 
by extrapolating the branches of Fig. 17 to higher values of 
$l_\sigma$. The physical origin of all these types of states 
(localized, sparse and extended) was explained above, and their 
appearance can be decoded by examining the structure of the 
"channel representation" of the Hamiltonian matrix (Fig.2). 
A detailed inspection of this matrix (see also Eq. (26) of Ref. 
\cite{qrch}) shows that the coupling between unperturbed states 
depends strongly on the values of the index $m$, labeling the 
transversal modes of the flat billiard. An unperturbed state 
specified by a large value of $m$ (an $m$ close to $M_{max}=62$) 
couples strongly to several other unperturbed states. In contrast,
a state with $m=1$ has practically no coupling to other states. 
In particular, the extremely localized states, corresponding to the 
first position on the left line of each brach of Fig. 17 occur 
because of the negligible coupling of the diagonal elements of the 
$H_{l,l'}$ matrix with $m=1$, the states on the second position of 
each branch occur for $m=3$, and so on with $m$ odd. Similarly, 
states on the right side of the branches result from elements of the 
$H_{l,l'}$ matrix with even values of $m$.

This structure is expected to prevail at all energy ranges since in 
any sufficiently large range of energies there are unperturbed states 
with all values of $m$ in $[1,M_{max}]$. Even deep in the 
semiclassical regime, extremely localized and sparse states will appear 
but less and less frequently since the energy differences between states 
of the same type increases with energy. Consequently, the strong 
fluctuations appearing in the SES and LDOS distributions (Figs. 11 
and 12) will tend to dissapear as $\hbar_{eff} \rightarrow 0$.

In the definitions of SES and LDOS, the averages were performed over 
the appropriate range of energy in order to take into account once all the
various types of states (Figs. 10 and 11 were calculated this way).

We can understand the origin of this type of regular structure by
analyzing the Hamiltonian matrix $H_{l,l'}$, see Fig. 2. A
detailed inspection shows that for the unperturbed states
(diagonal matrix elements), the coupling with other unperturbed
states depends on the value of the index $m$ which labels the
number of transversal modes in the billiard [see Eqs. (11-12)]. An
unperturbed state specified by a large $m$ (an $m$ close to
$M_{max}=62$, for any $n$) shows a strong coupling to several
other unperturbed states. In contrast, a state with $m=1$ (for any
$n$) shows practically no coupling to other states.

We can see then that the extremely localized states, corresponding
to the first position on the left hand side of each set of points in Fig. 16
occur because of the negligible coupling of the diagonal elements
of the $H_{l,l'}$ matrix with $m=1$. Similarly, the states on the
second position (on the left hand side) occur when $m=3$, 
and so on with odd $m$. States
on the right hand side of each set of points result from elements of the
$H_{l,l'}$ matrix with even values of $m$. This structure prevails
for all energies due to the existence of unperturbed $m=1$ states.
However, the intervals between the occurence of $m=1$ unperturbed
states increase as the energy increases. Even as the semiclassical
limit is approached, extremely localized states will appear,
except less and less frequent. Consequently the strong
fluctuations appearing in the SES and LDOS distributions (Figs. 10
and 11) will tend to dissapear.

\section{Concluding Remarks}

The quantum-classical correspondence for the chaotic motion of an 
electron in a periodic billiard was analyzed in terms of the classical 
analogues of the structure of eigenstates (SES) and local density of 
states (LDOS).

To construct the classical counterparts of the LDOS and SES, we 
first changed ({\it via} a canonical transformation) to some new 
curvilinear coordinates, where the new Hamiltonian incorporated the 
effects of the boundary into an effective interaction potential. 
Then the original system of a free particle colliding within a 2D 
rippled channel becomes equivalent to a 1D model of two interacting 
"particles", identified with the new coordinates. This procedure 
allowed us to study the chaotic properties of this system, using 
tools developed recently to analyze the role that interactions 
between particles play in the onset of quantum chaos. This example 
is quite instructive since it shows the relevance of ``one-body chaos", 
which is due to an external potential (or boundary conditions), to 
``two-body chaos", which results entirely from the interaction 
between particles. So far, these two mechanisms for producing chaos 
have been treated completely independently.

Using this approach, we found that in the case of strong chaos, the 
classical analogs of SES and LDOS agreed remarkably well with the 
global shapes of the quantum quantities. This correspondence enabled 
us to explain the main features of the SES and LDOS in terms of the
underlying classical motion of the system. Specifically, we found 
that classical unstable and stable periodic orbits of period-one give 
rise to the two pronounced side-band peaks observed in the SES and LDOS. 

On the other hand, we found that the statistical properties of 
eigenstates and individual LDOS  differ qualitatively from those
prescribed by the standard random matrix theory. Namely, in the case 
of strong chaos and deep in the semiclassical region (high energy), 
one expects the components of individual eigenstates to fluctuate in 
a statistically independent way around the mean, the envelope of the 
SES in the unperturbed energy basis. This expectation was confirmed 
when studying the structure of eigenstates of complex atoms and 
nuclei in the mean-field basis \cite{ce,zele}. In contrast, we have 
found that in our present  model the deviations turn out to be
extremely strong and not fully statistically independent. \\

In connection with this, a detailed analysis revealed quite 
unexpected regularities in localization measures (such as the inverse 
participation ratio) characterizing the eigenstates and individual
LDOS. We remark that these regularities do not disappear as energy 
increases. In particular, even for high energies, one can find
eigenstates that are strongly localized in the unperturbed energy 
representation ({\it i.e.}, slightly perturbed plane waves). This 
occurs because energy is not the only semiclassical parameter in a 
2D electron waveguide. Clearly, two unperturbed states of similar 
(or equal) energy values but different transverse mode numbers will 
react differently to a small perturbation.That is, in contrast to 
high-mode unperturbed states, the transverse wavelength $\Lambda_y$ 
for low-mode states can be many times larger than the amplitude of 
the perturbation and hence will remain essentially unperturbed. \\

Thus, for any value of energy (equivalently, for any small value of 
$\hbar$), there are {\it non-ergodic states}. It should be stressed 
that here we are discussing the ergodicity in the energy shell, which 
is determined by the classical motion of a particle in terms of the 
classical SES. Since the width of the energy shell is always finite 
due to the finite range of the interaction, the ergodicity of an 
eigenstate means that this eigenstate fills the whole energy shell 
with random (Gaussian) fluctuations of its components around the 
smooth energy dependence defined by the classical SES.

Naturally, with an increase of the energy, the {\it relative}
number of localized (in the energy shell) eigenstates tends to zero.
In this sense, there is no contradiction with the onset of quantum
ergodicity in the classical limit. However, this limit turns out to 
be achieved very slowly. Therefore, we remark that from a physical 
point of view, it is important to further the study of the statistical 
properties of eigenstates in the deep semiclassical regime, which may 
be different from those in the strictly classical limit ($\hbar=0$, 
a mathematical concept).

\section{Acknowledgments}
We acknowledge support from CONACyT (Mexico) Grant No. 26163-E

\newpage

\begin{center}
{\Large \bf
FIGURE CAPTIONS}
\end{center}
\vspace{.5cm}

\begin{description}

\item{\bf FIG. 1} Geometry of a rippled billiard.

\item{\bf FIG. 2} Central part of the $4030 \times 4030$ Hamiltonian matrix
$H_{l,l'}(k)$: $N_{max}=32$, and $M_{max}=62$. The $62 \times 62$
blocks corresponding to $n,n'=0, \pm 1, \pm 2, \pm 3$ are shown.

\item{\bf FIG. 3} (a) Unperturbed energy spectra as a function of $l$. (b)
Unperturbed energy spectra ordered in energy as a function of $l_{new}$.

\item{\bf FIG. 4} Example of eigenfunctions for different ways of odering
the unperturbed basis. (a) the state $\alpha = 2122$ as a
function of $l$; (b) the same state as a function of $l_{new}$.

\item{\bf FIG. 5} Left upper part of the matrix $w_l^\alpha$ corresponding to
$N_{max}=32$, and $M_{max}=62$.

\item{\bf FIG. 6} a) Energy of the unperturbed Hamiltonian $E^0(t)$ as a
function of time (in arbitrary units) for $E=1$. Note the
intermittent structure (from a noise-like to a stable-like) in
$E^0(t)$. b) Classical distribution $W_{cl}(E^0 \mid E)$
constructed from $E^0(t)$.

\item{\bf FIG. 7} Energy $E^0$, and $P_u^2$, $P_v^2$, $v$ as a function of time
(in arbitrary units) on a short time scale, after the substitution
of a trajectory $\phi(t) \in H$ onto $H^0$ for $E$ = 1.

\item{\bf FIG. 8} The same as in Fig. 7 for a different time interval.

\item{\bf FIG. 9} a) Energy $E(t)$ of the full Hamiltonian $H$ as a function
of time (in arbitrary units) after the substitution of 36
trajectories $\phi^0(t) \in H^0$ onto $H$ for $E^0$ = 1. b)
Corresponding classical LDOS $\omega_{cl}(E \mid E^0)$ constructed
from $E(t)$. Note that each trajectory contributes in a different
way to the classical distribution as can be seen in Fig. 9a, for
details see the text.

\item{\bf FIG. 10} Individual eigenfunctions $w_l^\alpha$ in the
energy representation and the classical distribution $W_{cl}(E^0
\mid E)$ (thick curve) as a function of the energy $ E^0_l/E^\alpha$
for $\alpha$= a) 111, b) 712, c) 1362, and c) 2002. The effective Planck's 
constant $\hbar_{eff}$ is a)0.027, b) 0.011, c) 0.008, and d) 0.006, 
respectively. The entropy localization length (defined in Sect. VIII) 
$l_{\cal H}$ is a) 23.6, b) 158.9, c) 483.6, and d) 709.

\item{\bf FIG. 11} Structure of eigenstates (SES) in the energy representation
$W(E^0_l \mid E^\alpha)$ and its classical counterpart $W_{cl}(E^0
\mid E)$ (thick curve). The average for the SES is taken over the
interval $1880<\alpha<2025$.

\item{\bf FIG. 12} LDOS $\omega(E^\alpha \mid E^0_l)$ and its classical counterpart
$\omega_{cl}(E \mid E^0)$ (thick curve). For the LDOS the average
is taken over the interval $1900<l<2000$.

\item{\bf FIG. 13} Typical high energy eigenfunctions for a) $\alpha$=1952,
b) $\alpha$=1953, and c) $\alpha$=1954. The localization measures
(defined below) are: a) $l_H$=745.9, $l_{ipr}$=529.7; b) $l_H$=232.9,
$l_{ipr}$=45.4; and c) $l_H$=2.7, $l_{ipr}$=3.2.

\item{\bf FIG. 14} Left: a localized and an extended eigenfucntions in the
basis representation for (a) $\alpha$ = 2082 (extremely localized
state with corresponding $m=2$), (b) $\alpha$ = 2083. Right: the
same eigenfunctions in configuration representations.

\item{\bf FIG. 15} Localization measures for eigenfunctions in dependence on
$\alpha$ for exact $\alpha^{th}$ eigenstates. (a) Entropy localization
length $l_{\cal H}$, (b) $l_{ipr}$ defined through the inverse participation
ratio, (c) the mean square root $l_\sigma$, and (d) the centroid $n_c$.

\item{\bf FIG. 16} Localization measures for the LDOS with $l$ standing for the
unperturbed state $|l>$. Same localization measures as in Fig. 15
are presented: (a) $l_{\cal H}$, (b) $l_{ipr}$, (c) $l_\sigma$, and (d) 
$n_c$.

\item{\bf FIG. 17} Detail of Fig. 16c for the range $1450 < l < 2100$. 
The numbers indicate specific states which are shown in
Fig. 18. The states are chosen from 3 different sets of states,
specified by almost the same values of $l_{\sigma}$.

\item{\bf FIG. 18} Structure of individual LDOS marked in Fig. 17. One can
see the similarity in the structure of the LDOS taken from
different sets, but with the same values of $l_{\sigma}$. 

\end{description}


\begin{thebibliography}{99}
\bibitem{note2} In Ref. \cite{qrch}, a slightly different transformation
was used. There $v \equiv y/(d + a \cos (x))$, which contains
the two parameters $a$ and $d$ but the wave function vanishes at
$y=1$.
\bibitem{NOTE} It is interesting to note that
(through the use of commutation
relations, like $[\hat P_{\alpha},g^{-1/4}]=-i\hbar
\partial_{\alpha} g^{-1/4}$], the
quantum Hamiltonian (8) becomes $H= \frac{1}{2}P_{\alpha}g^{\alpha
\beta}P_{\beta} + \hbar^2 Q(u,v)$,
where $Q(u,v)= (\epsilon/4) \left(
\xi_{uu}/g^{1/2}-3 \epsilon \xi_u ^2/g\right)$.
By comparison with the classical Hamiltonian (17), the quantity
$Q$ may be interpreted (see e.g., \cite{Dewitt2}) as a sort of
quantum potential that must be added to the covariant classical
expression in order to produce the correct quantum covariant
Hamiltonian.
\end{thebibliography}
\end{document}